# FOOTBALL: A NAÏVE APPROXIMATION TO THE EFFECT OF INCREASING GOAL SIZE ON THE NUMBER OF GOALS

J. Mira

(fajmirap@usc.es)

Departamento de Física Aplicada

Universidade de Santiago de Compostela

E-15782 Santiago de Compostela, Spain

## Abstract

The effect of increasing goal mouth size on the number of goals scored in a football match is discussed in a very preliminary and simple way, considering elastic collisions of the ball with the posts. The result is obtained on the basis of data taken from the Spanish Professional League, that show a high number of shots-to-post. Surprisingly, there is a direct correlation of the increase in goal mouth area with the increase of goals.

Football is the most popular sport in Europe, and a major area of economic activity: in Spain, for example, figures from the Professional Football League suggest that the football industry accounts for 1% of Gross National Product. In the rest of the world, football's influence is increasing apace. Given such significant repercussions, it is interesting to consider whether it might be possible to improve the game's entertainment value on the basis of analyses of play situations [1], or even by introducing changes in the rules.

Clearly, goals are the crowning moments of a match. Typically however, the number of goals scored is small, and not surprisingly many matches end as 0-0 draws. One proposal for increasing the number of goals has been to increase the size of the goal mouth. This proposal is being reinforced after Germany's World Cup 2006, where the average number of goals, 2.29, was the second worst in the history of the tournament. In fact, FIFA's President Joseph Blatter has declared that this could be the formula to make football "attractive again" by increasing its attacking character (take into account that highly defensive teams like Italy and France arrived to the final).

In an attempt to quantify the likely consequences of such a measure, summary statistics for First Division matches in the Spanish Professional League, 2000-2001 season [2], have been analysed. The total number of shots-at-goal hitting the post (and not going in) was 231. Given that the total number of goals in this period was 1095, in 380 matches (2.88 goals per match), expressing shots-to-post as a percentage of goals scored gives 21.1%, a higher proportion than expected in principle given the small frontal area of the posts (which are at most 12 cm thick).

The effect of varying goal size on the number of goals can be estimated as follows (under the simplifying assumptions that all shots are perpendicular to the goal line and that the ball's collisions with the cylindrical posts are perfectly elastic): if the goal mouth were expanded laterally and vertically by one ball-diameter (rules require ball diameter to be between 21.6 and 22.3 cm) plus 7 cm, then all current shots-to-post would be goals (whether without hitting the new posts or after hitting them). The 7-cm increase is necessary to ensure that the ball changes direction by at least 90 degrees after the collision, even in the least favourable case.

Of course, further refinements of the model might be considered (relating to shot angles, shot probabilities across the goal area, and new play situations created after rebounds from the posts of the enlarged goal mouth); however, the simple model is probably sufficient. Increasing goal size in such way means an increase from 2.44 m x 7.32 m (17.86 $m^2$) to 2.73 m x 7.90 m (21.57 $m^2$). This percentage increase in area, 20.8%, is almost equal to the above-noted increase in goals (21.1%), which is rather surprising. Taking into account that the increase in goals would occur at the edges of the goal, less accessible to the goalkeeper, in principle the probability of goal in the "new" area should be higher than in the central areas where the keeper spends more of his time.